\documentclass{ws-procs10x7}
\usepackage{amsmath}
\usepackage{amssymb}
\usepackage{graphicx}
\usepackage{array}
\newcommand{\beq}{\begin{equation}}
\newcommand{\eeq}{\end{equation}}
\newcommand{\bea}{\begin{eqnarray*}}
\newcommand{\eea}{\end{eqnarray*}}
\newcommand{\bei}{\begin{itemize}}
\newcommand{\eei}{\end{itemize}}

\def\dsl{\,\raise-.02ex\hbox{/}\mkern-11.5mu D}

\def\OMIT#1{{}}

\newcommand\spur{\raise.15ex\hbox{/}\kern-.57em } 

\newcommand\Dsl{\,\raise-.02ex\hbox{/}\mkern-11.5mu D} %this one can be subscripted

\newcommand{\CO}{{\cal O}}

\newcommand{\Vub}{|V_{ub}|}

\begin{document}

\title{%Accurate $\Vub$ from exclusive $B$ and $D$ decays.}
Theory of the endpoint region of exclusive rare B decays
}
\author{Benjamin Grinstein}

\address{University of California, San Diego; La Jolla, CA,
  92093-0319; USA\\E-mail: bgrinstein@ucsd.edu}

\twocolumn[\maketitle\abstract{In the ratio of ratios of rates for
radiative and semileptonic decays of $B$ and $D$ to $K^*$ and $\rho$ mesons the
non-computable form factors mostly cancel out. The resulting method
for the determination of $\Vub$  is largely free of hadronic
uncertainties, raising the prospect of a precise determination of 
$\Vub$. }]

\section{Motivation and Results}%1

%\subsection{subsection title}\label{subsec:one}%1.1
The third row of the CKM matrix is poorly determined because it
involves the top quark. Of the first two rows, the $V_{ub}$ element is
the least well known, with a precision of about 20\% in its magnitude,
$\Vub$. 

The measurement of $\Vub$ through inclusive charmless semileptonic $B$
decays is complicated by the large background from charm-full
semileptonic $B$ decays. Experimentally, cuts may be imposed to
suppress or eliminate the background, but this introduces theoretical
uncertainties. Some years ago it was guessed\cite{bauerguess} that
this method would eventually reach a precision of 5\%, but by now new
complications from theory have been discovered\cite{voloshin-stewart},
so the eventual precision of the method is expected to be
significantly worse.

Alternatively one may look to determine $\Vub$ from exclusive
semileptonic $B$ decays, such as $B\to\pi\ell\nu$ or $B\to\rho\ell\nu$. The difficulty
here is from theory: the form factors are not well known. Eventually,
unquenched lattice QCD calculations may produce accurate form factors
(at least in a restricted region of phase space, but that's all that
is needed to extract $\Vub$). However, the precision at the moment is
no better than\cite{hashimoto} 20\%.

In this talk I describe a method that relies on exclusive decays which
can be used to determine $\Vub$ to high precision, possibly a couple
of percent. The trick is to let experiment, rather than theorists,
determine the form factors. Hence, 
%
%\subsection{Results}\label{subsec:results}
the main result of our
work\cite{preciseVub,Grinstein:2002rm,Grinstein:2002cz}% is a recipe
can be framed as a recipe
for experiment. Measure the ratio of $B$ decay rates,
\begin{multline}
\label{eq:result}
\frac{{\mbox{d}\Gamma(\bar B\to \rho e\nu)}/{\mbox{d}q^2}}
{{\mbox{d}\Gamma(\bar B\to K^* \ell^+ \ell^-)}/{\mbox{d}q^2}}
= { \frac{|V_{ub}|^2}{|V_{tb} V_{ts}^*|^2}} \cdot\\
 \frac{8\pi^2}{\alpha^2}\cdot
\frac{1}{N_{\rm eff}(q^2)}
\frac{\sum_\lambda |H_\lambda^{B\to\rho}(q^2)|^2}
{\sum_\lambda |H_\lambda^{B\to K^*}(q^2)|^2}
\end{multline}
Here $q^2$ is the invariant mass of the lepton pair and the
measurement must be done at large $q^2$ (the energy of the 
$\rho$ or $K^*$ in the $B$ rest-frame must not exceed about 500~MeV).
The factor $N_{\rm eff}(q^2)$ is provided by theory. The ratio of
helicity amplitudes,
\beq
\label{eq:RB}
R_{B}(y) \equiv \frac{\sum_\lambda |H_\lambda^{B\to\rho}(y)|^2}
{\sum_\lambda |H_\lambda^{B\to K^*}(y)|^2} 
\eeq
can be obtained experimentally, as follows. Measure decays spectra for
$D\to \rho\ell\nu $ and $D\to K^*\ell\nu $, and from it construct the ratio
\beq
\label{eq:RD}
R_{D}(y) \equiv  \frac{\sum_\lambda |H_\lambda^{D\to\rho}(y)|^2}{\sum_\lambda |H_\lambda^{D\to K^*}(y)|^2}
\eeq
Once the ratios in (\ref{eq:RB}) and (\ref{eq:RD}) are expressed in terms of
$y=E_V/m_V$ ($V=\rho, K^*$), one can safely replace one for the other, 
\beq
R_{B}(y) =R_{D}(y)\Big(1 + {\cal O}(m_s(\frac{1}{m_c}-\frac{1}{m_b}))\Big) 
\eeq

In summary, to determine $\Vub$, measure the ratio of rates in
Eq.~(\ref{eq:result}), infer the ratio $R_{B}(y)$ from semileptonic
$D$ decays, and combine with the theory-provided $N_{\rm
  eff}(q^2)$. At NLL order, $N_{\rm  eff}(q^2_{\text{max}})=32.76$, varying by 2\%
as the renormalization scale $\mu$ is varied between 2.4~GeV and 9.6~GeV.
The $q^2$ dependence is mild (and fully known).

\begin{table*}[t]
\caption{Comparison of theoretical predictions of the  double-ratio $R_1$ of
decay constants of  heavy mesons \label{tab:R1}}
\begin{tabular}{ccccc}
 \hline \hline &Method&${f_{B_s}}/{f_{B_d}}$&${f_{D_s}}/{f_{D_d}}$
&$R_1$\\ \hline
&Relativistic Quark
 Model\cite{debert}&1.10$\pm$0.21&1.09$\pm$0.22&1.01$\pm$0.40\\  
&Quenched Lattice\cite{cbernard}&1.16(1)(2)(2)($^{+4}_{-0}$)&
1.14(1)($^{+2}_{-3}$)(3)(1)&1.02(2)(4)(4)($^{+4}_{-1}$)\\
&Unquenched Lattice\cite{tonogi} &&&1.018$\pm$0.006$\pm$0.010\\
&Quenched Lattice\cite{ryan}&1.15$\pm$0.03&1.12$\pm$0.02&1.03$\pm$0.05\\
&Unquenched Lattice\cite{ryan}&1.16$\pm$0.05&1.12$\pm$0.04&1.04$\pm$0.08\\
&QSR\cite{narison}&1.16$\pm$0.05&1.15$\pm$0.04&1.01$\pm$0.08\\ 
&RSM\cite{cvetic}&1.10$\pm$0.01&1.08$\pm$0.01&1.02$\pm$0.02\\
\hline\hline
\end{tabular}
\end{table*}

\section{Theory}
\OMIT{The rest of the talk we describe the theory underlying the claims
above.} 
\subsection{Double ratios}
Since the $K^*$ and the $\rho$ mesons are similar, one could hope that
the hadronic uncertainties would largely cancel in ratios. For
example, consider radiative decays:
\begin{equation}
\label{eq:badratio}
\frac{\Gamma (B\to K^*\gamma)}{\Gamma (B\to \rho \gamma)} 
\propto
           \left|\frac{F_{ B\to K^*}(0)}{F_{B\to \rho }(0)}\right|^2~.
\end{equation}
The proportionality factor, which includes calculable phase space, CKM
and short distance QCD corrections, has been omitted so that we may
focus our attention on the hadronic form factors, $F=F(q^2)$, which
are the main culprits for theoretical uncertainties. In the
flavor-$SU(3)$ limit \( F_{ B\to K^*}(0)=F_{B\to \rho }(0)\) so the
ratio in (\ref{eq:badratio}) is unity. However, flavor-$SU(3)$
symmetry is good to $\sim30\%$.  One may try to improve this situation
by guessing that the ratio of form factor is similar to the known
ratio of decay constants, \( F_{ B\to K^*}(0)/F_{B\to \rho }(0)\approx
f_K/f_\pi\), but there is no way of assessing precisely the error
incurred, so this is not what we want for precision physics.

If we are willing to measure more quantities we can do better. The idea
is to construct a ratio of quantities that is fixed to unity by two distinct
symmetries. Consider, for example, heavy meson decay
constants. In the $SU(3)$ flavor symmetry limit, two ratios are set
to unity
\beq
\label{eq:ratio1}
\frac{f_{B_s}}{f_{B}}=1\qquad{\rm and}\qquad \frac{f_{D_s}}{f_{D}}=1
\eeq
Similarly,  Heavy Quark Flavor Symmetry fixes different ratios of the
same quantities, 
\beq
\label{eq:ratio2}
\frac{f_{B_s}}{f_{D_s}}=\sqrt{\frac{m_c}{m_b}}\qquad{\rm and}\qquad  
\frac{f_{B}}{f_{D}}=\sqrt{\frac{m_c}{m_b}}
\eeq
The ratio of ratios can be written as the ratio of
the two ratios in either (\ref{eq:ratio1}) or (\ref{eq:ratio2}), so it
is protected from deviating from~1 by the two symmetries:\cite{doubleratios} 
\beq 
R_1= \frac{f_{B_s}/f_{B}}{f_{D_s}/f_{D}}= 1 
     + {\cal
       O}\left(m_s\left(\frac{1}{m_c}-\frac{1}{m_b}\right)\right)
\nonumber
\eeq

Table \ref{tab:R1} demonstrates how well this works. The ratio $R_1$
is computed by different theoretical methods, so the individual decay
constants and ratios  differ significantly between methods. But
because all methods incorporate flavor and heavy quark symmetry the
double ratio deviates from unity by a couple of percent in all
cases. Similarly estimates of double ratios of form factors
in semileptonic decays, $B\to D^{(*)}\ell\nu$\cite{boyd1},
$B\to(K^*,\rho )\ell\nu$\cite{lsw} and
$B\to(K,\pi)\ell\nu$\cite{boyd2}, give deviations from unity of a few
percent.

\subsection{Double ratio in $B\to\rho\ell\nu$}
Ligeti and Wise\cite{lig-wise} studied the use of double ratios in the
determination of form factors in $B\to \rho\ell\nu $. By measuring
form factors in $B\to K^*\ell^+\ell^-$, $D\to K^*\ell\nu$ and
$D\to\rho\ell\nu$ one can use a double ratio to determine
\beq
f^{(B\to\rho)}\approx f^{(B\to K^*)}\, \frac{f^{(D\to\rho)}}{f^{(D\to K^*)}}
 ~. 
\eeq

There are three stumbling blocks to complete this program:\cite{preciseVub}
\begin{enumerate}
\item The rates depend on several form factors, but the double ratio
relates individual form factors. Measuring individual form factors
places additional demands on experimental measurements.
\item There are form factors in $B\to K^*\ell^+\ell^- $ that
are not present in the semileptonic decays.  
\item There are long distance contributions to $B\to K^*\ell^+\ell^-$
(mostly from $B\to K^* (c\bar c)$ with $(c\bar c)\to\ell^+\ell^-
$).
\end{enumerate}
We address each problem in turn.

\begin{figure}
\centering
\includegraphics[height=2.4cm]{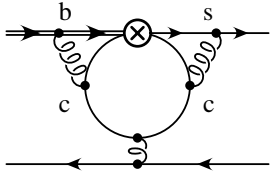}
\caption{Example of non-factorizable diagram contributing to the
  ``long-distance'' amplitude for $B\to
K^*\ell^+\ell^- $}
\label{fig:non-factLD}
\end{figure}

The solution to problem \#1 above is trivial. Let's ignore for now the
additional form factors from tensor operators that enter $B\to
K^*\ell\ell$ (difficulty \#2, above). Consider the semileptonic decay
rate  
\beq
\hspace{-0.8ex}\frac{d\Gamma(\bar B\to \rho e \nu)}{dq^2} = 
\frac{G_F^2 |V_{ub}|^2}{96 \pi^3 m_B^2}
q^2 |\vec q|\!\!\! \sum_{\lambda = \pm ,0}\!\!\! |H_\lambda|^2
\eeq
Here $H_\lambda$, $\lambda=\pm,0$ are helicity amplitudes. The point
is that the double ratio technique can be applied directly to the sums
$\sum_{\lambda = \pm ,0} |H_\lambda|^2$. If we were comparing
$(B\to\rho\ell\nu)/(B\to K^*\ell\nu)$ to $(D\to\rho\ell\nu)/(D\to K^*\ell\nu)$ 
we would have accomplished our goal. The problem is that there is no
$B\to K^*\ell\nu$ decay. So we consider $B\to K^*\ell\ell$ instead,
but this introduces the additional complications \#2 and \#3.

To address these issues, recall that the effective $\Delta B=-\Delta
S=1$ Hamiltonian,
\bea
{\cal H}_{\rm eff} = -\frac{G_F}{\sqrt2} V_{tb} V_{ts}^* \sum_{i=1}^{10} C_i(\mu)
Q_i(\mu)\,,
\eea
has four-quark operators $Q_1$--$Q_6$, {\it e.g.,}
\(Q_2  = (\bar s c)_{V-A} (\bar c b)_{V-A}\),  a transition magnetic  moment
operator,
\beq
Q_7 = \frac{e}{8\pi^2} m_b \bar s_\alpha \sigma_{\mu\nu} (1+\gamma_5)
b_\alpha F_{\mu\nu},
\eeq
an analogous color moment operator $Q_8$, and vector and axial current
operators 
\begin{equation}
Q_{9,10} = \frac{e^2}{8\pi^2}(\bar sb)_{V-A} (\bar e e)_{V,A}
\end{equation}
The Wilson coefficients $C_{9,10}$ are  larger than the rest. Roughly, at
$\mu=m_b$ one has $C_{9,10}\approx5$, $C_2\approx1$,
$C_{1,7}\approx1/3$, and the rest much smaller. This is good news,
since the matrix elements of $Q_{9,10}$ are given in terms of the
same form factors as for the semileptonic decay.

The rate for $B\to K^* \ell\ell$ depends on the tensor form factors
from the operator $Q_7$. These form factors are given in
terms of the vector and axial form factors of the semileptonic decays
up to $1/m_Q$ corrections by heavy quark symmetry, provided $y\sim1$
(recall $y=E_V/m_V$). One may write the rate in terms of helicity
amplitudes
\begin{multline}
\frac{d\Gamma(B\to K^* e^+ e^-)}{dq^2} = 
\frac{4G_F^2 |V_{tb} V_{ts}^*|^2 \alpha^2}{3m_B^2 (4\pi)^5}\times\\
q^2 |\vec q\,| \sum_{\lambda = \pm 1,0}
\left\{ |H_\lambda^{(V)}|^2 + |H_\lambda^{(A)}|^2 \right\}\,,
\end{multline}
and the form factor relations give
\beq H_\lambda^{(V)}(q^2) = C_9\Big(1 + \delta(q^2) +
{\CO}(\Lambda/m_b)\Big) H_\lambda(q^2),%\nonumber
\eeq
where $\delta(q^2) = ({2m_b^2}/{q^2}) ({C_7}/{C_9}) $. It is not the
smallness of the correction $\delta(q^2)$ that matters, since it is
computable. Rather, it is that the corrections to the form factor
relations, denoted by ``${\CO}(\Lambda/m_b)$'', have an additional
suppression of $C_7/C_9\sim10^{-1}$ allowing for high precision in the
method.

Finally we have to deal with the contributions of the operators
$Q_1$--$Q_6$. These are the long-distance effects of difficulty \#3
listed above. Note, however, that the use of form factor relations
has forced us to work at $q^2$ near
$q^2_{\text{max}}=(M_B-M_{K^*})^2$. This is well 
above the threshold
for charm pair production. Since $q^2$ is large we can perform an
operator product expansion (OPE) in inverse powers of $q^2$. To avoid
positive powers of $m_b^2$ it is best to take $\sqrt{q^2}\sim m_b$ and
expand in both large scales simultaneously. The ``long-distance
effect'' is replaced by a sum of local terms that can be truncated given
a desired accuracy. 

\begin{figure}
\centering
\includegraphics[height=6.3cm]{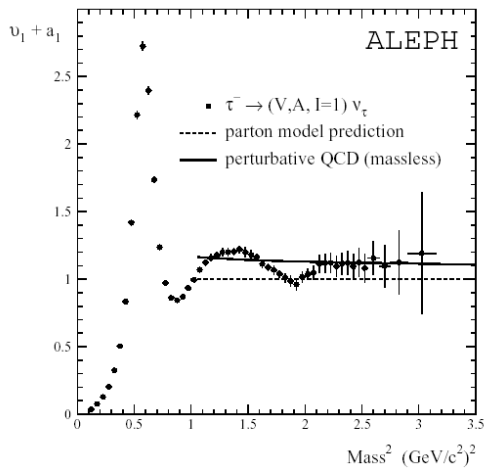}
\caption{ALEPH's\protect\cite{Barate:1998uf} charged current form factor from $\tau$
  decay. Local quark-hadron duality is a poor approximation in the
  resonant region but quickly improves a few GeV$^2$ above that.}
\label{fig:QH-duality}
\end{figure}

The good news is that the leading terms in the expansion give rise to
computable amplitudes: they are expressed in terms of the same form
factors that appear in semileptonic decays. The bad news is that this
OPE is performed in the time-like region. It is applicable only to the
extent that quark-hadron duality is a good
approximation. Fig.~\ref{fig:QH-duality} shows how quark hadron
duality works in one example, the form factor for the charged current
obtained from $\tau$ decay data. The bad news is not so bad: since the
``long-distance'' effect is a small contribution to the $B\to
K^*\ell\ell$ rate, one can tolerate a 50\% error in quark-hadron
duality without spoiling the  determination of $\Vub$ to a few
percent accuracy. Notice also that no assumption of factorization was
made.

\section*{Acknowledgments}
Partial support is
from the DOE under Grant DE-FG03-97ER40546.

\end{document}